\documentstyle[aps,epsf]{revtex}
\begin{document}
\draft
\preprint{ }
\title{Theory of coherent time-dependent transport in one-dimensional
multiband semiconductor superlattices}
\author{Jon Rotvig}
\address{Geophysical Department, University of Copenhagen,
Juliane Maries Vej 30, DK-2100 Copenhagen {\O}, Denmark}
\author{Antti-Pekka Jauho}
\address{Mikroelektronik Centret, Technical University of Denmark,
Bldg. 345east, DK-2800 Lyngby, Denmark}
\author{Henrik Smith}
\address{{\O}rsted Laboratory, H. C. {\O}rsted Institute,
Universitetsparken 5, University of Copenhagen, DK-2100 Copenhagen {\O},
Denmark}
\date{\today}
\maketitle
\begin{abstract}
We present an analytical study of
one-dimensional semiconductor superlattices in
external electric fields, which may be time-dependent.
A number of general results for the (quasi)energies and
eigenstates are derived.  An equation of motion for the
density matrix is obtained for a two-band model, and the
properties of the solutions are analyzed.
An expression for the current is obtained.  Finally,
Zener-tunneling in a two-band tight-binding model is considered.
The present work gives the background and
an extension of the theoretical
framework underlying our recent Letter [J. Rotvig
{\it et al.}, Phys. Rev. Lett. {\bf 74}, 1831 (1995)],
where a set of numerical simulations were presented.
\end{abstract}
\section{Introduction}
Studies of Bloch electrons under the influence of external electric
fields attract at present intensive theoretical attention
(some recent papers can be found in Refs.
\onlinecite{Holthaus1,Holthaus2,Holthaus3,Ignatov1,Ignatov2,Ignatov3,Zhao1,Zhao2,Zhao3,Meier1,Meier2,Emin,Hart});
this is a natural consequence of recent
experimental advances, which include the
observation of Bloch oscillations \cite{Waschke},
and studies of photon-assisted transport \cite{Guimaraes,Keay1,Keay2}.
Thus the classic predictions of Bloch\cite{Bloch} and
Zener\cite{Zener} have finally been verified, and
the arena is open for new investigations and ideas.
The physics of superlattices in external fields is extremely
rich due to the large number of parameters that can
be controlled quite freely.  Many
of the physical properties are  sensitive
functions of these parameters, and hence slight
adjustments in their values
allow one to move between different physical regimes,
both experimentally and theoretically.
Examples of such parameters are the miniband structure
of the superlattice (which can be controlled by varying the
composition and/or thickness of the layers comprising the
superlattice to meet the requirements of the
particular investigation) or the intensities and frequencies of
the external fields. This flexibility allows
one to study many different physical phenomena. The
 papers quoted above have
addressed such varying physical phenomena
as (i) dynamical localization or band-collapse
(originally discussed in \onlinecite{Dunlap}) either in
dc or ac electric fields, (ii) interplay of field-induced localization
and Anderson localization due to disorder,
(iii) chaotic motion of charge carriers,
or (iv) full-scale numerical integration of semiconductor Bloch
equations, which allow one to study interaction
effects, such as exciton dynamics.

Much of the interest has been caused by the
need of understanding the interplay of Bloch oscillations
({\it i.e.}, coherent time-periodic motion of charge carriers
in one band),
Zener tunneling between Bloch bands, and interaction
effects.
We studied recently
\cite{letter} a two-band tight-binding model, and
found via a direct numerical solution of the density-matrix
equation-of-motion a rather complicated temporal behavior for the diagonal
(in miniband index) component. This structure could, however,
be interpreted in terms of intuitive physical arguments.  Our
numerical study was facilitated by a substantial amount of  analytic
background work, and the purpose of this paper is to provide a
full account of the formal developments.
In subsequent sections we
shall derive a number of formal results, which we believe
to be new, or generalizations of previously known results,
and we give mathematical proofs of several statements made
in our Letter.  We also give a detailed discussion of
the density-matrix equation-of-motion, upon which
our previously reported numerical work was based, and which can
be used as a starting point for future studies, for example
consideration of relaxation or loss of phase coherence
due to scattering.

The paper is organized as follows.
In Section II we derive the energy spectrum for a $N$-band superlattice,
in static and time-dependent fields, and apply the general
results to the tight-binding two-band superlattice of
Ref.\onlinecite{letter}.  Section III is devoted to
the equation-of-motion analysis of the density matrix,
and a derivation of an expression for the current,
from which semiclassical intraband and quantum-mechanical
tunneling contributions can be identified.  We also give
some analytic results for Zener tunneling, which we used
in the interpretation  of our numerical
results of Ref.\onlinecite{letter}.  The final
section gives our conclusions.
\section{ Energy spectra}
\label{sec1}

In this section we consider the energy spectrum of an electron in a
superlattice in the presence of a static or time-periodic electric field,
which causes coupling between the bands. Our formulation allows us to
obtain a  simple proof of the existence of the Wannier Stark (WS) ladders in
an arbitrary superlattice with band coupling, when a static field is applied.
The number of interpenetrating WS ladders is equal to the number of
zero-field energy bands in the model. This was first proven by Emin and Hart
\cite{Emin} by splitting the electric potential into two parts with a sawtooth
and a staircase shape; our proof does not rely on this particular
construction.
In the final part of this section
we introduce the tight-binding model of a superlattice
containing  two minibands used in the simulations of \cite{letter},
and establish a connection to the general results derived in
the beginning of this section.

\subsection{Electronic motion in a superlattice}

We consider a semiconductor
superlattice with a growth direction parallel to
the $x$-axis. The lattice period is denoted by $d$ and is repeated $N_p=L/d$
times, where $L$ is the length of the superlattice.
In the effective mass approximation \cite{Bastard} the
wave function satisfies the usual Schr\"odinger equation with  the
scalar-potential Hamiltonian $H_{\phi}$ given by
\begin{equation}
H_\phi(x,t)=\frac{p_x^2}{2m}+V_c(x)+eE(t)x\;,
\label{Hscalpot}
\end{equation}
where $m$ is the effective mass, $V_c(x)$ the periodic potential with
period $d$, and ${\bf E}(t)=(E(t),0,0)$ is the electric field. For $E(t)=0$,
the eigenstate problem $H_\phi(x)\psi(x)=\epsilon\psi(x)$ is solved by the
usual Bloch states (BS)
\begin{equation}
\phi_{nK}(x)=\frac{1}{\sqrt{L}}e^{iKx}u_{nK}(x)
\end{equation}
with eigenvalues
$\epsilon_n(K)$ associated with the miniband indexed by $n$.
In a finite electrical field we follow \cite{Krieger1}
and introduce accelerated Bloch states (ABS) when calculating
the energy spectrum. We assume that the field is applied at $t=0$ and
define the vector potential ${\bf A}(t)=(A(t),0,0)$ by
$A(t)=-\int_0^t E(t')\,dt'$. Then the semiclassical time evolvement
of the crystal momentum
${\bf k}(t)=(k(t),0,0)$ is given by
\begin{equation}
k(t)=K-(e/\hbar)\int_0^t E(t')\,dt'=K+eA(t)/\hbar\\;.
\end{equation}
At this point it is useful to introduce the Hamiltonian $H_{A}$
in which the electrical field is represented by the vector potential
instead of the scalar potential
\begin{equation}
H_A(x,t)=\frac{[p_x+eA(t)]^2}{2m}+V_c(x)\;.
\label{Hvecpot}
\end{equation}
The ABS, defined as
\begin{equation}
\psi_{nK}(x,t)=\frac{1}{\sqrt{L}}e^{iKx}u_{nk(t)}(x),
\end{equation}
are instantaneous eigenstates to the time-dependent
Hamiltonian $H_A(x,t)$,
\begin{equation}
\label{HA_ABS}
H_A(x,t)\psi_{nK}(x,t)=\epsilon_n[k(t)]\psi_{nK}(x,t)\;.
\end{equation}
The set of ABS forms an orthonormal basis. The so-called
Houston function \cite{Houston}  $\phi_{nk(t)}(x)$ can
be expressed in
the ABS basis as
\begin{eqnarray}
\label{aBS}
\phi_{nk(t)}(x)&=&e^{ieA(t)x/\hbar}\psi_{nK}(x,t)=
e^{ieA(t)x/\hbar}\frac{1}{\sqrt{L}}e^{iKx}u_{nk(t)}(x)\nonumber\\
&=&\frac{1}{\sqrt{L}}e^{ik(t)x}u_{nk(t)}(x)\;.
\end{eqnarray}
The Houston function is thus
an ordinary BS corresponding to the wavevector ${\bf k}(t)$.

We shall write solutions to the Schr\"odinger equation,
\begin{equation}
\label{Sch1}
H_\phi(x,t)\psi(x,t)=i\hbar\frac{\partial\psi(x,t)}{\partial t}\;,
\end{equation}
in the form $\psi(x,t)=e^{-i\epsilon t/\hbar}u(x,t)$, where $u(x,t)$
is spatially periodic with period $L$. We thus have
\begin{equation}
\label{Sch}
H_\phi(x,t)u(x,t)=\epsilon u(x,t)+i\hbar\frac{\partial u(x,t)}{\partial t}\;.
\end{equation}
For a static electric field the function
$u$ can be chosen to be independent of time and
$\epsilon$ is the energy. If the field is periodic in time with a period
$T_{ac}$, Floquet's theorem states that  $u$ will have the same
periodicity. The Floquet solutions define the quasienergy $\epsilon$ which
is a generalization of the energy.

Now let us  expand $\psi(x,t)$ in the ABS according to
\begin{equation}
\label{expan}
\psi(x,t)=\sum_{nK}C_{nK}(t)e^{ieA(t)x/\hbar}\psi_{nK}(x,t)\;.
\end{equation}
We then insert (\ref{expan}) into (\ref{Sch1}) and use
\begin{equation}\label{eq11}
i\hbar{\partial\psi_{nK}(x,t)\over\partial t} =
i\hbar {e^{iKx}\over\sqrt{L}}\nabla_k u_{nk(t)}(x){\dot k}\;,
\end{equation}
where we defined $\dot{k}\equiv\partial k/\partial t$.
Since $\nabla_k u_{nk}$ is a periodic function  with period
$d$, it may be expanded in terms of the functions
$u_{n'k}$ according to
\begin{equation}\label{eq12}
i\nabla_k u_{nk(t)}=  \sum_{n'} u_{n'k(t)} R_{n'n}[k(t)]\;,
\end{equation}
where the expansion coefficients $R_{n'n}$ are the matrix elements
\begin{equation}
R_{n'n}(k)
=\frac{i}{d}\int_{-d/2}^{d/2}dx\,u^\ast_{n'k}(x)\nabla_k u_{nk}(x)\;.
\end{equation}
Combining Eqs.(\ref{eq11}-\ref{eq12})
with ${\dot k} = -eE/\hbar$
results in the following coupled equations
for the
expansion coefficients $C_{nK}(t)$:
\begin{equation}
\label{final}
i\hbar\frac{\partial C_{nK}(t)}{\partial t}=\epsilon_{n}[k(t)]C_{nK}(t)
-F(t)\sum_{n'}R_{nn'}[k(t)]C_{n'K}(t)\;,
\end{equation}
where $F(t) = -eE(t)$.
Equations similar to Eq.(\ref{final}) were derived by
Krieger and Iafrate\cite{Krieger1}. Our formulation differs from theirs, in
that they use a  gauge transformation in order to eliminate the
spatially non-periodic term $eE(t)x$. We avoid the gauge transformation
and thereby keep the notion of energy. Instead we introduce a common factor $\exp[ieA(t)x/\hbar]$ in the ABS-expansion of the wavefunction. As seen from Eq.(\ref{expan}) this is the same as using an ordinary BS-expansion of the wavefunction:
\begin{eqnarray}
\label{expan2}
\psi(x,t)&=&\sum_{nK}C_{nK}(t)\phi_{nk(t)}(x)\nonumber\\
&=&e^{-i\epsilon t/\hbar}\sum_{nK}c_{nK}(t)\phi_{nK}(x)\;,
\end{eqnarray}
where $c_{nK}(t)=C_{n,K-eA(t)/\hbar}(t)e^{i\epsilon t/\hbar}$. The function
$u$ is found from Eq.(\ref{expan2})
\begin{equation}
\label{u}
u(x,t)=\sum_{nK}c_{nK}(t)\phi_{nK}(x)\;.
\end{equation}

The  set  of coupled equations (\ref{final}) for the
coefficients $C_{nK}(t)$ will now
be shown to lead to the existence of
WS-ladders in static electric fields.

\subsubsection{Static fields}

In the following we consider a model periodic potential corresponding to
a finite number $N$ of energy bands and demonstrate the
existence of WS-ladders in static fields.
For a static field, $F(t)=F(>0)$, the $c_{nK}$'s in Eq.(\ref{u}) are
independent of time. From this condition and the periodicity of the
$c_{nK}$'s in $K$ it follows that
\begin{equation}
\label{dcdirectcond}
C_{nK}(t+T_B)=C_{nK}(t)e^{-i\epsilon T_B/\hbar}\;,
\end{equation}
where $T_B=2\pi/\omega_B$ and $\omega_B=Fd/\hbar$.
For each $K$ in the Brillouin zone (BZ) there are $N$ linearly independent solutions ${\bf C}_{Kj}(t)$ ;
$j=1,...,N$ to the system Eq.(\ref{final}). From Eq.(\ref{dcdirectcond}) we
see that the energy for a specific solution ${\bf C}_{Kj}(t)$ is only
determined modulo $\hbar\omega_B$. However,  the substitution
$\epsilon\rightarrow\epsilon+p\hbar\omega_B\;,\;p={\rm integer}$, yields
${\bf c}_{Kj}(t)\rightarrow{\bf c}_{Kj}(t)e^{ip\omega_Bt}$.
Only one of the corresponding solutions ${\bf c}_{Kj}^p(t)$,
$p={\rm integer}$, can therefore be time-independent. Let us
denote the corresponding energy by $\epsilon_{Kj}$. The system of equations
(\ref{final})
has periodic coefficients with period $T_B$. By applying the $N$-dimensional
Floquet theorem we know that the solutions are of the form
${\bf C}_{Kj}(t)=e^{i\omega_{Kj}t}{\bf P}_{Kj}^0(t)$ , where $\omega_{Kj}$ is
only determined modulo $\omega_B$ and
${\bf P}_{Kj}^0(t+T_B)={\bf P}_{Kj}^0(t)$. From Eq.(\ref{dcdirectcond}) the
energies of the Floquet solution
${\bf C}_{Kj}(t)$ are $\epsilon_{Kj}^p=-\hbar\omega_{Kj}+p\hbar\omega_B$ ,
$p={\rm integer}$. Let $p(K,j)$ be given by
$\epsilon_{Kj}^{p(K,j)}=\epsilon_{Kj}$. Then the energy spectrum is
\begin{equation}
\label{spectrum0}
\epsilon_{Kj}=
-\hbar\omega_{Kj}+p(K,j)\hbar\omega_B\;;\;j=1,...,N\;;\;K\in BZ\;.
\end{equation}

We can obtain more information by eliminating the $K$ dependence in
Eq.(\ref{final}): the semiclassical motion of $k(t)$ is monotonic and
can therefore be inverted to yield
$t(k)=\hbar(k-K)/F$. By use of the definition
 $D_n(k)=c_{nk}e^{-i\epsilon k/F}$ we obtain
 $C_{nK}(t)=D_n[k(t)]e^{i\epsilon K/F}$. Eq.(\ref{final}) is then
reduced to a system of equations which is independent of the initial momentum
$K$,
\begin{equation}
\label{static}
iF\frac{\partial D_n(k)}{\partial k}=\epsilon_n(k)D_n(k)
-F\sum_{n'}R_{nn'}(k)D_{n'}(k)\;.
\end{equation}
A direct calculation shows that
\begin{equation}
\label{dccond}
D_n(k+2\pi/d)=D_n(k)e^{-i\epsilon T_B/\hbar}\;.
\end{equation}
As before the energy is only determined modulo $\hbar\omega_B$. If
for a definite solution ${\bf D}_j(k)$ to Eq.(\ref{static}) we make the replacement
$\epsilon\rightarrow\epsilon+p\hbar\omega_B\;,\;p={\rm integer}$,
then ${\bf c}_{Kj}\rightarrow{\bf c}_{Kj}e^{iKpd}$. Therefore the displaced
energy corresponds to a spatial
translation of the solution by $-pd$, {\it i.e.},
the eigenstates
in terms of the ${\bf D}_j(k)$'s are only determined within a direct lattice
displacement. The system Eq.(\ref{static}) has periodic coefficients with
period $2\pi/d$. Appealing again to the $N$-dimensional Floquet theorem we
conclude that
the solutions are of the form
${\bf D}_j(k)=e^{ikr_j}{\bf P}_j^1(k)\;;\;j=1,...,N$. The Floquet functions
${\bf P}_j^1(k)$ are periodic with period $2\pi/d$. By calculating
${\bf D}_j(k+2\pi/d)$ and using Eq.(\ref{dccond}) we therefore obtain the
energy spectrum
\begin{equation}
\label{spectrum1}
\epsilon_{jp}=-Fr_j+p\hbar\omega_B\;;\;j=1,...,N\;;\;p\;{\rm integer}\;.
\end{equation}
We see that the spectrum can be visualized as $N$ WS energy ladders. The
relative positions of these are expressed in terms of the Floquet
coefficients $r_j$. This completes our general proof
of the existence of Wannier Stark (WS) ladders in
a static electric field.

It should be noted that the $\omega_{Kj}$'s and $r_j$'s are connected.
This can be demonstrated as follows. Floquet solutions for ${\bf D}(k)$ can
also be found by using the Floquet solutions for ${\bf C}_K(t)$.
For an arbitrary choice of $K$ we put
${\bf D}_j(k)=e^{-i\epsilon K/F}{\bf C}_{Kj}(k)
=e^{i\hbar\omega_{Kj}k/F}e^{-i(\epsilon+\hbar\omega_{Kj})K/F}
{\bf P}_{Kj}^0(k)$. The energy spectrum is then
$\epsilon_{Kj}^p=-\hbar\omega_{Kj}+p\hbar\omega_B,\;p\;{\rm integer}$.
Uniqueness of the spectrum implies that the energies
$\{\hbar\omega_{Kj}\;|\;j=1,...,N\}$ are equal to the WS ladder positions
$\{Fr_j\;|\;j=1,...,N\}$ except for individual multiples of $\hbar\omega_B$.

\subsubsection{Time-periodic fields}

In this section we consider
time-periodic solutions $u(x,t)$ to Eq.(\ref{Sch}).
We can derive general results concerning
the quasienergy spectrum using the properties of the
${\bf C}_{K}$-coefficients.  As we shall see,
the interplay between
the spatial periodicity and the temporal periodicity
can have significant consequences for the quasienergy
spectrum.

Specifically, let us consider
a time-dependent electric field with  period
$T_{ac}$. Since the coefficients $c_{nK}(t)$ in Eq.(\ref{u}) are periodic
in $t$ and  $A(t+T_{ac})=A(t)+A(T_{ac})$, we find that
\begin{equation}
\label{condperiodic}
C_{nK}(t+T_{ac})=C_{n,K+\Delta K}(t)e^{-i\epsilon T_{ac}/\hbar}\;,
\end{equation}
where $\Delta K=eA(T_{ac})/\hbar$. The quasienergy  $\epsilon$
for a given solution ${\bf C}_{Kj}(t)$ to Eq.(\ref{final}) is only given
modulo $\hbar\omega_{ac}$, where $\omega_{ac}=2\pi/T_{ac}$. The substitution
$\epsilon\rightarrow\epsilon+\hbar\omega_{ac}l\;,\;l={\rm integer}$,
results in ${\bf c}_{Kj}(t)\rightarrow{\bf c}_{Kj}(t)e^{i\omega_{ac}t}$.
Thus the Floquet state changes only by a time-dependent phase factor.

Let us now consider two special cases for $\Delta K$.  First,
if $\Delta K=0$ modulo $2\pi/d$, as for a harmonic field, the
condition Eq.(\ref{condperiodic}) is equivalent to Eq.(\ref{dcdirectcond})
in the static case. All momenta in the $BZ$ are then independent.

A special case arises if
$\Delta K=\frac{2\pi}{d}\frac{p}{q}$ with $p$ and $q$ integers\cite{Zhao1}.
Now a finite number of crystal momenta couple and the quasienergies are
determined from
\begin{equation}
\label{accond}
C_{nK}(t+qT_{ac})=C_{nK}(t)e^{-i\epsilon qT_{ac}/\hbar}\;.
\end{equation}
The quasienergies
 are now only defined modulo $\hbar\omega_{ac}/q$. The Brillouin zone
collapses to $[-\pi/qd,\pi/qd]$ and  the quasienergy
spectrum consists of fractional quasienergy ladders.

We have so far assumed that the system stays in equilibrium until
$t=0$, when it is instantaneously coupled to the electric field.
This is obviously an idealization, and it is natural to ask whether a finite
switch-on period will change the general results.  We have repeated the above
analysis, but allowed for a finite switch-on period $[0,T_0]$.
The upshot is that all formal results for the quasienergy
spectrum still apply, however, one must change the initial momentum label
$K$. The new momentum values, $K\to K-eA(T_0)/\hbar$, correspond to a
translation in reciprocal space \cite{Proof1}.

\subsection{Tight-binding model}

In the rest of this paper we consider a superlattice model with only two
bands,  $n = a,b$.  The Hamiltonian is
\begin{eqnarray}\label{tbhamilt}
H&=&\sum_l\Bigl\{\bigl[\Delta_0^a-F(t)X^a-F(t)ld\big]a_l^\dagger a_l
+\bigl[\Delta_0^b-F(t)X^b-F(t)ld\bigr]b_l^\dagger b_l\nonumber\\
&&-\frac{\Delta_1^a}{4}(a_{l+1}^\dagger a_l+a_l^\dagger a_{l+1})
+\frac{\Delta_1^b}{4}(b_{l+1}^\dagger b_l+b_l^\dagger b_{l+1})
-F(t)X^{ab}(a_l^\dagger b_l+b_l^\dagger a_l)\Bigr\}\;.
\end{eqnarray}
The terms involving the constants $X^{a}$ and $ X^{b}$  describe
a site-independent shift, due to the electric field, of
the unperturbed  energies  $\Delta_0^{a}$
and $\Delta_0^{b}$, while the band coupling
is described by the term involving
the constant $X^{ab}$.
 The remaining terms in the Hamiltonian
involve the site-dependent potential energy $-F(t)ld$
and the hopping between nearest-neighbor sites.
Due to  the terms involving $X^a$ and $X^b$
the model represents a generalization of the two-band
model which was first solved in \cite{Fukuyama}.
The reason for introducing these parameters is that
we can construct, as shown below, a mapping between the tight-binding model
and a general two-band model (when the band couplings
are described with the parameters $R^a$, $R^b$, and $R^{ab}$).
The various energy parameters describing the two-band
superlattice are summarized in Fig.\ \ref{fig1}.

We shall first consider static electric fields.
Then the Schr\"odinger equation, $H_\phi u=\epsilon u$, in the
site-representation is solved  by making the ansatz
$u=\sum_l(u^a_l a_l^\dagger+u^b_l b_l^\dagger)|0\rangle$,
and projecting out the $l$:th component,
which results
in the following coupled equations for the expansion coefficients:
\begin{eqnarray}
\label{tbdceqn}
(\Delta_0^a-FX^a-Fld)u^a_l-\frac{\Delta_1^a}{4}(u^a_{l-1}+u^a_{l+1})
-FX^{ab}u^b_l&=&\epsilon u^a_l\nonumber\\
(\Delta_0^b-FX^b-Fld)u^b_l+\frac{\Delta_1^b}{4}(u^b_{l-1}+u^b_{l+1})
-FX^{ab}u^a_l&=&\epsilon u^b_l\;.
\end{eqnarray}
For vanishing field the spectrum consists of two bands
\begin{equation}
\label{cosinebands}
\epsilon^{a,b}(K)=\Delta_0^{a,b}\mp(\Delta_1^{a,b}/2)\cos(Kd)\;,
\end{equation}
corresponding to the plane-wave solutions $u^a_l=e^{iKld}\;,\;u^b_l=0$ and
$u^a_l=0\;,\;u^b_l=e^{iKld}$, respectively. When the field is finite,
we depart from the procedure
used in \cite{Fukuyama} by applying the discrete Fourier transformation
\begin{eqnarray}
u^\nu_l&=&\sum_K u^\nu_K e^{iKld}\nonumber\\
u^\nu_K&=&\frac{1}{N_p}\sum_l u^\nu_l e^{-iKld}\;,
\end{eqnarray}
directly on Eq.(\ref{tbdceqn}). With the notation
$u^\nu_K=P^\nu_K e^{i\epsilon K/F}$ the result is
\begin{equation}
\label{tbdceqnf}
iF\frac{\partial}{\partial K}
\left({\renewcommand{\arraystretch}{1.5}\begin{array}{c}
P^a_K\\P^b_K
\end{array}}\right)
=\left({\renewcommand{\arraystretch}{1.5}\begin{array}{cc}
\epsilon^a(K)-FX^a&-FX^{ab}\\-FX^{ab}&\epsilon^b(K)-FX^b
\end{array}}\right)
\left({\renewcommand{\arraystretch}{1.5}\begin{array}{c}
P^a_K\\P^b_K
\end{array}}\right)\;,
\end{equation}
and
\begin{equation}
\label{tbdccond}
P^\nu_{K+2\pi/d}=P^\nu_Ke^{-i\epsilon T_B/\hbar}\;.
\end{equation}
The form of Eq.(\ref{tbdccond}) is analogous to Eq.(\ref{dccond}).
Furthermore, Eq.(\ref{tbdceqnf}) in the tight-binding model has the form of
Eq.(\ref{static}) if the bands are identical. In addition, the band couplings
in the general model must be time-independent and satisfy $X^a=R^a$,
$X^b=R^b$ and $X^{ab}=R^{ab}=R^{ba}$. The last equation requires that
$X^{ab}$ is real. If these conditions are fulfilled, the tight-binding model
describes exactly the energy spectrum Eq.(\ref{spectrum1}). The condition
$X^a-R^a=X^b-R^b=0$ can be relaxed in the case where only the difference
spectrum between the two WS ladders is required to be exact. Then it is sufficient that $X^a-R^a=X^b-R^b=\Delta R$ \cite{Proof2}.

We now return to the solution of Eq.(\ref{tbdceqnf}). The diagonal
matrix elements
in this equation can be eliminated by transforming
$P^\nu_K=\widetilde{P^\nu_K}
e^{-\frac{i}{F}\int_0^K(\epsilon^\nu-FX^\nu)\,dK'}$.
One then finds \cite{Fukuyama}
$\widetilde{P^a_K}$
\begin{equation}
\label{secondorder}
\frac{\partial^2\widetilde{P^a_K}}{\partial K^2}
+i\frac{\hbar\widetilde{\omega}_d}{F}
\frac{\partial\widetilde{P^a_K}}{\partial K}
+\left(X^{ab}\right)^2\widetilde{P^a_K}=0\;,
\end{equation}
where  we defined $\widetilde{\omega}_d(K)=\omega_d(K)+FX_-/\hbar$. Here
the band separation is given by
$\hbar\omega_d(K)=\epsilon^b(K)-\epsilon^a(K)$, and we also introduced
$X_-=X^a-X^b$. The
boundary condition is
\begin{equation}
\widetilde{P^a_K}_{+2\pi/d}=\widetilde{P^a_K}
e^{-i(\epsilon-\Delta_0^a+FX^a)T_B/\hbar}\;.
\end{equation}
Let us next consider
a part $I$ of the Brillouin zone, where  $\omega_d(K)$ is a
weak function of $K$ (i.e., the bands are almost flat).  Specifically,
we demand that
$\Delta\omega_d(I) = {\rm max}[\omega_d(K)]-{\rm min}[\omega_d(K)]$,
$K\in I$, is
small compared to $\langle\omega_d\rangle_I+FX_-/\hbar$ (the average is
calculated over $I$).  In this case Eq.(\ref{secondorder}) is easily solved,
and one finds
\begin{equation}
\label{partialsolutions}
\widetilde{P^a_K}=e^{-i\hbar (\tilde{\omega}_d\pm\omega_l)K/2F}\;.
\end{equation}
Here the  frequency $\omega_l$ is given
 by $\omega^2_l=\omega_c^2+\widetilde{\omega}_d^2$
with $\omega_c=2(|X^{ab}|/d)\omega_B$. Let us define the
mean band separation $\omega_0=\langle\omega_d\rangle_{BZ}$ and
$\widetilde{\omega}_0=\omega_0+FX_-/\hbar$. The integration interval can be
extended to the whole Brillouin zone when
$\Delta\omega_d(BZ)\ll\widetilde{\omega}_0$. The solutions
Eq.(\ref{partialsolutions}) then correspond to the spectrum
\begin{eqnarray}
\label{flatbandspectrum}
\epsilon^\pm&=&\frac{\Delta_0^a+\Delta_0^b}{2}-F\frac{X_+}{2}
\pm\frac{\hbar\Omega_l}{2}+p\hbar\omega_B\;,\;p\;{\rm integer}\;,\nonumber\\
\Omega_l^2&=&\omega_c^2+\widetilde{\omega}_0^2\;,
\end{eqnarray}
where $X_+=X^a+X^b$. Thus
the two interpenetrating WS ladders are positioned symmetrically around the
band midpoint, translated by $-FX_+/2$, and the ladder separation is
$\hbar\Omega_l$. The exact form of the spectrum (\ref{flatbandspectrum})
can also be obtained in a much more direct way by looking for localized
solutions to Eq.(\ref{tbdceqn}). We restrict the solution to be confined
within the $p$:th unit cell. Solving
\begin{equation}
\left|{\renewcommand{\arraystretch}{1.5}\begin{array}{cc}
\Delta_0^a-F(X^a+pd)-\epsilon&-FX^{ab}\\-FX^{ab}&\Delta_0^b-F(X^b+pd)-\epsilon
\end{array}}\right|=0\;,
\end{equation}
we find the spectrum (\ref{flatbandspectrum}). The difference spectrum
obtained under the flat band (or localization) condition will reappear in
the following section when we discuss the electronic response in a static
field in the case of nearly flat bands.

Next let us turn to time-periodic fields, when the right hand side of
Eq.(\ref{tbdceqn}) acquires the additional terms
$i\hbar\frac{\partial}{\partial t}u^{a(b)}_l$.
Following Zhao et al. \cite{Zhao3} we substitute
\begin{equation}
\label{Zhaotransf}
u^\nu_l=Q^\nu_l \exp\{i[\epsilon t+eA(t)ld]/\hbar\}\;,
\end{equation}
and Fourier transform, whence
\begin{equation}
\label{tbaceqnf}
i\hbar\frac{\partial}{\partial t}
\left({\renewcommand{\arraystretch}{1.5}\begin{array}{c}
Q^a_K\\Q^b_K
\end{array}}\right)
=\left({\renewcommand{\arraystretch}{1.5}\begin{array}{cc}
\epsilon^a[k(t)]-F(t)X^a&-F(t)X^{ab}\\-F(t)X^{ab}&\epsilon^b[k(t)]-F(t)X^b
\end{array}}\right)
\left({\renewcommand{\arraystretch}{1.5}\begin{array}{c}
Q^a_K\\Q^b_K
\end{array}}\right)\;.
\end{equation}
>From the Fourier transform of (\ref{Zhaotransf}) we get the boundary condition
\begin{equation}
\label{tbaccond}
Q^\nu_K(t+T_{ac})=Q^\nu_{K+\Delta K}e^{-i\epsilon T_{ac}/\hbar}\;,
\end{equation}
with $\Delta K = eA(T_{ac})/\hbar$, as above.
A comparison of Eq.(\ref{tbaceqnf}) with Eq.(\ref{final}), and
Eq.(\ref{tbaccond}) with Eq.(\ref{condperiodic}), shows that in order to
obtain a correspondence we must set the Fourier variable $K$ equal to the
initial crystal momentum in the general model. The (difference) quasienergy
spectrum in the tight-binding model and the (difference) spectrum obtained
in the general model are then seen to be identical under the same conditions
as in the static case.

\section{Response}

We have thus far examined the general properties of the energy
spectrum and the wave functions of a superlattice in an external
field.  The next step is to compute the physical
quantities, such as densities or drift velocities.
To this end we must construct an appropriate kinetic
equation.  The simplest approach would be to consider
the Boltzmann
equation. However, since we are interested in the properties of
systems with two or more occupied minibands, where
a coherent
band-to-band
transfer plays an essential role, it is necessary to
use a method, which allows one to
consider quantum mechanical tunneling.
In the following sections we shall use the density matrix description.

\subsection{The density-matrix equations of motion}

In order to describe the coherent transport we shall start from
the equations of motions for the density matrix in the limit where
collisions may be neglected. Such equations of motions were derived
by Krieger and Iafrate \cite{Krieger2,Krieger3}. Thus we take as our
starting point Eq.(51) of Ref.\cite{Krieger2}, adapted to the case
of two minibands.
The density matrix is defined as
\begin{equation}
\rho_{nKn'K'}(t) = \langle\psi_{nK}(x,t)|\rho(x,t)|\psi_{n'K'}(x,t)\rangle\;,
\label{rhodef}
\end{equation}
and for coherent motion its diagonal elements ($\rho^a_K\equiv\rho_{aKaK}$
etc.) satisfy
\begin{eqnarray}
\label{diagonal_a}
\frac{\partial\rho^a_K(t)}{\partial t}
&=&\frac{1}{2}{\rm Re}\left\{h(K,t)\int_0^t h^\ast(K,t')
\left[\rho^b_K(t')-\rho^a_K(t')\right]\,dt'\right\}
\\
\label{diagonal_b}
\frac{\partial\rho^b_K(t)}{\partial t}
&=&\frac{1}{2}{\rm Re}\left\{h^\ast(K,t)\int_0^t h(K,t')
\left[\rho^a_K(t')-\rho^b_K(t')\right]\,dt'\right\}\;,
\end{eqnarray}
where
\begin{equation}
\label{h-func}
h(K,t)=-\frac{2F(t)R^{ab}[k(t)]}{\hbar}
e^{-i\int_0^t\left[\omega_d[k(t')]+F(t')R_-[k(t')]/\hbar\right]\,dt'}\;,
\end{equation}
and $R_-(k)=R^a(k)-R^b(k)$. Defining
$\rho_{\pm}(K,t)=\rho^a_K(t)\pm\rho^b_K(t)$ we get from
Eq.(\ref{diagonal_a},\ref{diagonal_b})
\begin{eqnarray}
\label{plus}
\dot{\rho}_+(K,t)&=&0\\
\label{minus}
\dot{\rho}_-(K,t)
&=&-{\rm Re}\left\{h(K,t)\int_0^t h^\ast(K,t')\rho_-(K,t')\,dt'\right\}\;,
\end{eqnarray}
where we defined $\dot{\rho}\equiv \partial\rho/\partial t$.
Eq.(\ref{plus}) is simply a statement of particle number conservation,
while the integro-differential equation (\ref{minus}) determines
the kinetic properties of our system.

As it stands, Eq.(\ref{minus}) is unsuited for
analytical applications, because of the non-Markovian
``collision term" (presently, the ``collisions" consist of band-to-band
transfers, dressed with the effects of the external fields).
However, it can be solved numerically by the technique described in
Ref.\cite{Cahlon}. We now prove that Eq.(\ref{minus}) can be reduced to a
third order differential equation, and then handled by standard numerical
techniques, if we assume that $u\neq0$ except in equilibrium. In addition
it will be shown that the form of this equation is invariant after a finite
switch-on period $[0,T_0]$. At the end of this time interval the response belonging to an initial momentum $K$ depends only on the position $k(T_0)$
of the semiclassical motion of $K$, and on the boundary conditions of
$\rho_-$ at $t=T_0$.

Since the different $K$-components do not
mix in (\ref{minus}), we can keep the initial momentum $K$ fixed. In order
to ease the notation
we write Eq.(\ref{h-func}) as $h=ue^{i\phi}$. Then the relation
$u\dot{h}=(\dot{u}+iu\dot{\phi})h$ holds. Using this, and differentiating
Eq.(\ref{minus}) we get
\begin{equation}
\label{helpeqn}
u\ddot{\rho}_-=\dot{u}\dot{\rho}_- -u\dot{\phi}
\,{\rm Re}\left\{ih\int_0^t h^\ast\rho_-\,dt'\right\}-u^3\rho_-\;.
\end{equation}
An auxiliary function $w$ is next defined by
$w=-{\rm Im}\left\{h\int_0^t h^\ast\rho_-\,dt'\right\}$. It
satisfies the equation
\begin{equation}
\label{helpeqn1}
u\dot{w}=\dot{u}w+u\dot{\phi}\dot{\rho}_-.
\end{equation}
We proceed by expressing $w$ in terms of another
auxiliary function, $w=u\widetilde{w}$. Using
$u\dot{\widetilde{w}}=\dot{\phi}\dot{\rho}_-$, we obtain
\begin{equation}
w(t)=u(t)\int_0^t\frac{\dot{\phi}(t')\dot{\rho}_-(t')}{u(t')}\,dt'.
\end{equation}
This allows one to rewrite Eq.(\ref{helpeqn}) as
\begin{equation}
\label{intdifeqnf}
\ddot{\rho}_- -\frac{\dot{u}}{u}\dot{\rho}_- +u^2\rho_-
+u\dot{\phi}\int_0^t\frac{\dot{\phi}\dot{\rho}_-}{u}\,dt'=0\;.
\end{equation}
Finally,
dividing Eq.(\ref{intdifeqnf}) by $u\dot{\phi}$, differentiating and
multiplying by $u^3\dot{\phi}^2$, we arrive at the third-order
differential equation
\begin{equation}
\label{thirdorder}
p_3\overdots{\rho}_-+p_2\ddot{\rho}_-+p_1\dot{\rho}_-+p_0\rho_-=0\;,
\end{equation}
where
\begin{eqnarray}
p_3&=&u^2\dot{\phi}\nonumber\\
p_2&=&-u(2u\dot{u}\dot{\phi}+u\ddot{\phi})\nonumber\\
p_1&=&(-u\ddot{u}+2\dot{u}^2+u^4+u^2\dot{\phi}^2)\dot{\phi}
+u\dot{u}\ddot{\phi}\nonumber\\
p_0&=&u^3(\dot{u}\dot{\phi}-u\ddot{\phi})\;.
\end{eqnarray}
>From Eq.(\ref{minus}) and Eq.(\ref{intdifeqnf}) above we get the boundary conditions $\dot{\rho}_-(K,0)=0$ and $\ddot{\rho}_-(K,0)=-u^2\rho_-(K,0)$.
These equations are readily numerically integrated,
and a few special cases were described in our recent Letter \cite{letter}.

\subsubsection{The current}

The results derived above can be used to obtain an expression for
the drift velocity $\bar{v}(t)$.  The result can be expressed
as a sum of a semiclassical, intraband term, and a term
with quantum mechanical origin, which incorporates Zener tunneling.

Omitting the details, we state our result:
\begin{equation}
\label{velocity1}
\bar{v}(t)=\sum_{\nu K}\rho^\nu_K(t) v^\nu_K(t)
+2\sum_K\left[{\rm Re}\,\rho^{ba}_K(t)\,{\rm Re}\,v^{ab}_K(t)
-{\rm Im}\,\rho^{ba}_K(t)\,{\rm Im}\,v^{ab}_K(t)\right]\;,
\end{equation}
where
\begin{eqnarray}
v^\nu_K(t)&=&\frac{1}{\hbar}\frac{\partial\epsilon^\nu[k(t)]}{\partial k}\\
v^{ab}_K(t)&=&-i\omega_d[k(t)]R^{ab}[k(t)]\;.
\end{eqnarray}
The non-diagonal density matrix elements
$\rho^{ba}_K$
can expressed in terms of the
diagonal elements,
\begin{equation}
u(t)\rho^{ba}_K(t)=-\frac{i}{2}h(K,t)\int_0^t h^\ast(K,t')\rho_-(K,t')\,dt'
\;.
\end{equation}
Using the auxiliary function $w$ defined in the previous section,
we obtain
\begin{eqnarray}
{\rm Re}\left\{u(t)\rho^{ba}_K(t)\right\}
&=&-\frac{u}{2}\int_0^t\frac{\dot{\phi}\dot{\rho}_-}{u}\,dt'\\
{\rm Im}\left\{u(t)\rho^{ba}_K(t)\right\}
&=&-\frac{1}{2}\,\dot{\rho}_-(K,t)\;.
\end{eqnarray}
If $R^{ab}$ is real, Eq.(\ref{velocity1}) can be further reduced, and
one finds
\begin{equation}
\label{velocity2}
\bar{v}(t)=\sum_{\nu K}\rho^\nu_K(t) v^\nu_K(t)
-\frac{1}{2F(t)}\sum_K\dot{\rho}_-(K,t)\hbar\omega_d[k(t)]\;.
\end{equation}
In Eq.(\ref{velocity1}) or (\ref{velocity2}) the first sum is the
semiclassical contribution to the drift velocity, while the second sum
accounts  for Zener tunneling.
Since $\rho^a_K(t)+\rho^b_K(t)$ = const. (determined by the initial
conditions), we can express Eq.(\ref{velocity2})
in terms of $\rho_-(K,t)$ alone.  Thus, a numerical solution of
the third-order differential equation derived in the previous
section,
Eq.(\ref{thirdorder}), is sufficient to determine the full time-dependence
of $\bar{v}(t)$.  This, however, must be done for each initially
occupied $K$-value, and we leave applications for future work.

\subsection{Zener tunneling}
The numerical calculations of Zener resonances based on (\ref{thirdorder})
were discussed in our recent Letter\cite{letter}.
Since tunneling between bands plays an important role in
the interpretation of our results we now consider the
condition for Zener tunneling in the present two-band model.
The electrical field is assumed  to be static $F(t)=F(>0)$.

\subsubsection{Analytical considerations}

\paragraph{Weak coupling limit.}
We shall show how a two-band version of Eq.(\ref{final}) can give
Zener resonances, following the perturbative method used by
Mullen et al., Ref.\cite{Mullen},
which is valid for weak band coupling. The two-band system is
\begin{equation}
\label{final_twoband}
i\hbar\frac{\partial}{\partial t}
\left({\renewcommand{\arraystretch}{1.5}\begin{array}{c}
C_K^a\\C_K^b
\end{array}}\right)
=\left({\renewcommand{\arraystretch}{1.5}\begin{array}{cc}
\epsilon^a-FR^a&-FR^{ab}\\-FR^{ba}&\epsilon^b-FR^b
\end{array}}\right)
\left({\renewcommand{\arraystretch}{1.5}\begin{array}{c}
C_K^a\\C_K^b
\end{array}}\right)\;.
\end{equation}
The system is linear and time-periodic with period $T_B$. The solution
$(C_K^a(NT_B),C_K^b(NT_B))^t$ can then be obtained from the initial state
$(C_K^a(0),C_K^b(0))^t$ by applying the time propagator ${\bf U}_0$ $N$
times on the initial state. The propagator  ${\bf U}_0$
is found as follows: By defining
$C_K^\nu(t)=\widetilde{C}_K^\nu(t)e^{-\frac{i}{\hbar}
\int_0^t(\epsilon^\nu-FR^\nu)\,dt'}$ the diagonal in Eq.(\ref{final_twoband})
is eliminated,
\begin{equation}
\frac{\partial}{\partial t}
\left({\renewcommand{\arraystretch}{1.5}\begin{array}{c}
\widetilde{C}_K^a\\\widetilde{C}_K^b
\end{array}}\right)
=\left({\renewcommand{\arraystretch}{1.5}\begin{array}{cc}
0&G^{ab}\\-(G^{ab})^\ast&0
\end{array}}\right)
\left({\renewcommand{\arraystretch}{1.5}\begin{array}{c}
\widetilde{C}_K^a\\\widetilde{C}_K^b
\end{array}}\right)\;,
\end{equation}
where $G^{ab}=i\frac{FR^{ab}}{\hbar}e^{-i\int_0^{T_B}\tilde{\omega}_d\,dt'}$.
We work in the limit of small band couplings, $|G^{ab}(t)T_B|\ll 1$.
To first order in $\Gamma^{ab}=\int_0^{T_B}G^{ab}\,dt'$ the initial state $(1,0)^t$ develops in a Bloch period $T_B$ to $(1,-(\Gamma^{ab})^\ast)^t$, and the orthogonal state $(0,1)$ to $(\Gamma^{ab},1)^t$. The time propagator is
then
\begin{equation}
{\bf U}_0
=e^{-\frac{i}{2\hbar}
\int_0^{T_B}(\epsilon^a+\epsilon^b-F(R^a+R^b))\,dt'}
\left({\renewcommand{\arraystretch}{1.5}\begin{array}{cc}
M&\Gamma^{ab}M\\
-(\Gamma^{ab}M)^\ast&M^\ast
\end{array}}\right)\;,
\end{equation}
where $M=e^{\frac{i}{2}\int_0^{T_B}\tilde{\omega}_d\,dt'}$.
According to ref.\cite{Mullen}
the resonance condition is that there should be
a phase difference $2\pi p$, $p$ integer, between the diagonal elements.
In our case, the resonance condition is
therefore $\int_0^{T_B}\tilde{\omega}_d\,dt'=2\pi p$ or
\begin{equation}
\label{Mullenres}
\frac{1}{Fd}
=\frac{1}{\hbar\omega_0}\left(p-\frac{\langle R_-\rangle}{d}\right) 
\;;\;p\;{\rm integer}\;.
\end{equation}

\paragraph{Static flat-band limit.}
Let us now discuss the case where the band coupling is not assumed to be
small. For a static field Eq.(\ref{thirdorder}) reduces to \cite{letter}
\begin{equation}
\label{dcthirdorder}
\dot{\phi}\overdots{\rho}_- -\ddot{\phi}\ddot{\rho}_-
+\dot{\phi}(u^2+\dot{\phi}^2)\dot{\rho}_- -u^2\ddot{\phi}\rho_- =0\;.
\end{equation}
We restrict ourselves to the case when
$\omega_1\equiv\omega_d-\omega_0\ll\widetilde{\omega}_0$.
It is furthermore assumed that the band couplings $R$ are independent of time
and we denote $X^a=R^a$, $X^b=R^b$ , and $X^{ab}=R^{ab}=R^{ba}$. We have
\begin{eqnarray}
u&=& 2(|X^{ab}|/d)\omega_B\equiv\omega_c\nonumber\\
\dot{\phi}&=&-(\widetilde{\omega}_0+\omega_1)\;.
\end{eqnarray}
Eq.(\ref{dcthirdorder}) is then
\begin{equation}
\label{dcthirdorder2}
-(\widetilde{\omega}_0+\omega_1)\overdots{\rho}_-
+\dot{\omega}_1\ddot{\rho}_-
-(\widetilde{\omega}_0+\omega_1)
\left(\omega_c^2+(\widetilde{\omega}_0+\omega_1)^2\right)\dot{\rho}_-
+\omega_c^2\dot{\omega}_1\rho_- =0\;.
\end{equation}
We write Eq.(\ref{dcthirdorder2}) in the form
\begin{equation}
(L_0+L_1+L_r)(\rho_0+\rho_1+\rho_r)=0\;,
\end{equation}
where $L_\mu$ and $\rho_\mu$ , $\mu=0,1,r$ , are of the zeroth, first, and
higher order in $\omega_1$. We get
\begin{eqnarray}
L_0\rho_-&=&-\widetilde{\omega}_0\overdots{\rho}_-
-\widetilde{\omega}_0\Omega_l^2\dot{\rho}_-\nonumber\\
L_1\rho_-&=&-\omega_1\overdots{\rho}_-
+\dot{\omega}_1\ddot{\rho}_-
-\omega_1(\Omega_l^2+2\widetilde{\omega}_0^2)\dot{\rho}_-
+\dot{\omega}_1\omega_c^2\rho_-\nonumber\\
L_r\rho_-&=&-\omega_1^2(3\widetilde{\omega}_0+\omega_1)\dot{\rho}_-\;.
\end{eqnarray}

The boundary conditions are $\rho_0(K,0)=\rho_-(K,0)$,
$\rho_1(K,0)=\rho_r(K,0)=0$, $\dot{\rho}_\mu(K,0)=0$, and
$\ddot{\rho}_\mu(K,0)=-\omega_c^2\rho_\mu(K,0)$. The zeroth order equation
$L_0\rho_0=0$ is readily integrated:
\begin{equation}
\label{rhozero}
\ddot{\rho}_0+\Omega_l^2\rho_0=\widetilde{\omega}_0^2\rho_0(0)\;.
\end{equation}
Using the zeroth order equation and Eq.(\ref{rhozero}) in the first order
equation $L_0\rho_1+L_1\rho_0=0$, and integrating, we find
\begin{equation}
\label{rhoone}
\ddot{\rho}_1+\Omega_l^2\rho_1=-\widetilde{\omega}_0
\left[\omega_1\left(\rho_0-\rho_0(0)\right)
+\int_0^t\omega_1\dot{\rho}_0\,dt'\right]\;.
\end{equation}
It is now clear that $\rho_0$ only contains the frequency $\Omega_l$. If
$\omega_1(K)$ is an analytic function then the spectrum of $\omega_1$ is
$p\omega_B$ , $p$ integer. From Eq.(\ref{rhoone}) we see that the Fourier
spectrum of $\rho_1$ consists of the frequencies $\Omega_l+p\omega_B$ , $p$
integer.
\par
Let us turn to the particular case with cosine bands, Eq.(\ref{cosinebands}).
As shown in the previous section the cosine bands can be generated by a
simple tight-binding model. For finite field values the energy spectrum
is given by Eq.(\ref{flatbandspectrum}).
Performing the integrals we obtain
\begin{eqnarray}\label{pertsol}
\rho_0(K,t)&=&\frac{\rho_0(K,0)}{\Omega_l^2}
\left(\widetilde{\omega}_0^2+\omega_c^2\cos\Omega_lt\right)\nonumber\\
\rho_1(K,t)&=&\frac{\widetilde{\omega}_0
(\Delta_1^a+\Delta_1^b)\omega_c^2\rho_0(K,0)}
{2\hbar\Omega_l^2(\Omega_l^2-\omega_B^2)\omega_B}
\Bigl\{\omega_B\cos Kd
+\omega_B\cos(\omega_B t+Kd)
+(\Omega_l-\omega_B)\cos Kd\cos\Omega_l t
-\Omega_l\cos(\Omega_l t+Kd)
\nonumber\\
&&\hspace{2.7cm}+\frac{\Omega_l-\omega_B}{2}
\cos\bigl[(\Omega_l+\omega_B)t+Kd\bigr]
-\frac{\Omega_l+\omega_B}{2}\cos\bigl[(\Omega_l-\omega_B)t-Kd\bigr]\Bigr\}\;,
\end{eqnarray}
where
$\Omega_l$ was defined in (\ref{flatbandspectrum}).
Thus to first order in $\omega_1$ the response contains the first
intra WS transition $\omega_B$ and the inter WS transitions
$\Omega_l\;,\;\Omega_l\pm\omega_B$.

The structure of the perturbative solution
(\ref{pertsol}) seems to suggest
that the response frequency spectrum always
corresponds to intra and inter transitions between (quasi) energy ladders.
This expectation is proven analytically in
Appendix A for static fields. The numerical
examples in \cite{letter} also support this conclusion for time-periodic
fields.

\subsubsection{A numerical example}

In our recent Letter \cite{letter} we presented
examples of a numerical
solution of the equation of motion for the density-matrix
$\rho_-(K=0,t)$.  It was found that at certain, relatively
sharp defined values of
the applied field the time-dependence of $\rho_-(0,t)$
exhibited a full inversion, i.e., its values ranged from
+1 (the initial value corresponding to a carrier in
band $a$) to -1 (carrier in band $b$).  The sharp resonances,
corresponding to complete band-to-band transfer, were termed
``Zener resonances".  These features were shown to
be intimately related to the Wannier-Stark ladder structure
of the system: the resonances occurred when the level differences
of the double ladder (corresponding to the two-band system)
was at minimum.  In the present section we describe the results
of extensive simulations which generalize our earlier results
in the following two respects:
(i) We allow the interband coupling, described by a parameter
$\xi=X^{ab}/d$, to vary from zero to large values; and
(ii) We examine the equation of motion for 41 different
$K$-points, covering the entire Brillouin zone.

The results of our numerical work, which consists of three
separate calculations, are summarized in Fig.\ \ref{fig2}.
The first calculation consists of an evaluation
of the local minima of {\it the normalized WS ladder
separation} as a function of field strength. These are represented by
diamonds in Fig.\ \ref{fig2}.
The normalized WS ladder separation is constructed
in three steps: (i) Compute the
differences of the energy levels
in the two WS ladders; (ii) Divide the differences
by $Fd$; and (iii) Select
the part of the normalized difference spectrum that
belongs to $[0,0.5]$.
It is seen from Fig.\ \ref{fig2} that by
increasing the interband coupling the local minima are shifted monotonously
towards higher field values and that they vanish in succession. The local
minima are numbered from the high field end at zero band coupling. The first
five local separation minima are depicted (when they exist) in
Fig.\ \ref{fig2}.

The second part of Fig.\ \ref{fig2} is a comparison of part one with the
local minima of the WS ladder separation in the case where the double
ladder is given by Eq.(\ref{flatbandspectrum}). The normalized difference
spectrum $\epsilon_{dif}=\pm(\epsilon^+-\epsilon^-)/Fd$ is
\begin{equation}
\epsilon_{dif}=q\pm\sqrt{4\xi^2+\left(\frac{\Delta^{ab}+FX_-}{Fd}\right)^2}
\;\;\;\;;\;q\;{\rm integer}\;.
\end{equation}
In our numerical example we have put $X^a=X^b=0$, thus $X_-=0$.
The normalized WS ladder separation have local minima (equal to
zero) whenever $q-\sqrt{4\xi^2+(\Delta^{ab}/Fd)^2}=0$ or
\begin{equation}
\frac{1}{Fd}=\frac{1}{\Delta^{ab}}\sqrt{q^2-4\xi^2}\;.
\end{equation}
In the limit of zero band coupling we get $1/Fd=q/\Delta^{ab}$ in
exact agreement with the correct result. In the high field limit the
local minima should be at $\xi=q/2$. As seen from Fig.\ \ref{fig2}, the
flat band (or localized) model tends to overshoot the exact findings,
but it provides an important hint about where to find the local minima.
In general the flat band model is not formally correct: the numerical
calculations show that the two WS ladders never coincide (level repulsion)
for non-zero band coupling. But even in our case, where the bands are far
from being flat, it estimates well the local minima.

The last part of Fig.\ \ref{fig2},
denoted by error bars, is a calculation of Zener resonances,
based on examining the time-development of $\rho_-(K,t)$.
As showed in our Letter\cite{letter} the number of oscillations in
$\rho_-(K,t)$ within a plateau increases when the field is decreased. A
plateau is better defined at the centre than at the terminal points where
the deviation from a constant value of $\rho_-(K,t)$ is largest. Thus for
Zener resonances with high index it is not sufficient to calculate
$\rho_-^{min}(K)$. Instead we define the centre of a plateau as the
time where the oscillation amplitude of $\rho_-(K,t)$ is at local minimum.
A Zener resonance is then a field value at which any of the plateau centre
values have reached $-1$. For a fixed interband coupling the first five
Zener resonances have been estimated (when they exist) for $41$ evenly
distributed $K$ points in the Brillouin zone. Some dispersion was detected,
and this is indicated by the height of the error bars in Fig.\ \ref{fig2}.
In the weak coupling limit we find zero dispersion and the Zener resonances
coincide with Eq.(\ref{Mullenres}).
As seen from Fig.\ \ref{fig2}, there is an excellent overall
correspondence between local minima of the normalized WS ladder
separation on one hand, and the Zener resonances on the other hand.

\section{Conclusion}

Our main results can be summarized as follows.
(i) The wave function for an electron in a periodic
potential, such as in a semiconductor superlattice,
under the influence of a (time-dependent)
uniform electric
field is expanded in terms of accelerated Bloch states.
The properties of the expansion coefficients allow us
to give a unified treatment of several different results,
obtained by a variety of techniques in the literature, for the
superlattice.
In static fields,
the existence of the Wannier-Stark ladder can be deduced, and in
the time-dependent case, with a periodic external field, a fractional
quasienergy spectrum emerges for certain values of the external
field.
(ii) A mapping is constructed between a two-band tight-binding
model, and the general system considered in (i). This allows
one to make several general statements concerning the properties
of the two-band system, both in static and time-dependent situations.
(iii) A density-matrix equation of motion is developed, and the
resulting third-order differential equation is shown to be well-suited
for both analytic and numerical studies of Zener tunneling in
the two-miniband system.  We also give new numerical results for
Zener resonances, identified in our earlier work \cite{letter},
and interpret them analytically applying the tools developed
in this work.

\appendix

\section{The response frequency spectrum}

We shall demonstrate that the response frequency spectrum in a dc field
consists of elements corresponding to intra- and inter-WS transitions.
In Sect.\ref{sec1} we found the linearly independent solutions
\begin{equation}
\psi_{Kj}(x,t)=e^{-i\epsilon_{Kj}t/\hbar}\sum_{n}c_{Kjn}\phi_{nK}(x)
\end{equation}
to the Schr\"odinger equation by using the $N$-dimensional Floquet theorem.
Now expanding a general solution as
\begin{equation}
\psi(x,t)=\sum_{Kj}\alpha_{Kj}\psi_{Kj}(x,t)\;,
\end{equation}
we obtain for the diagonal element $\rho_{nK}(t)$ of the density operator
$\rho(x,t)=|\psi(x,t)\rangle\langle\psi(x,t)|$:
\begin{eqnarray}
\label{rhodiasca}
\rho_{nK}(t)&=&\langle\psi_{nK}(x,t)|\rho(x,t)|\psi_{nK}(x,t)\rangle
\nonumber\\
&=&\sum_{\stackrel{K'j'n'}{K''j''n''}}
\alpha_{K'j'}\alpha_{K''j''}^\ast
c_{K'j'n'}c_{K''j''n''}^\ast
e^{-i(\epsilon_{K'j'}-\epsilon_{K''j''})t/\hbar}
\langle\psi_{nK}(x,t)|\phi_{n'K'}(x)\rangle
\langle\phi_{n''K''}(x)|\psi_{nK}(x,t)\rangle\;.
\end{eqnarray}
The ABS are time-periodic functions with period $T_B$. Thus $\rho_{nK}(t)$
contains only frequencies which are a subset of energy differences between
the WS ladders.

In Section III the gauge was changed. We now show that the response frequency
spectrum still consists of energy differences between the WS ladders for the
scalar-potential Hamiltonian. In the new gauge the Schr\"odinger equation is
on the form
\begin{equation}
H_A(x,t)\psi(x,t)=i\hbar\frac{\partial\psi(x,t)}{\partial t}\;.
\end{equation}
We expand the wave function directly in the ABS
$\psi(x,t)=\sum_{nK}B_{nK}(t)\psi_{nK}(x,t)$ and find the
system\cite{Krieger1}
\begin{equation}
i\hbar\frac{\partial B_{nK}(t)}{\partial t}=\epsilon_{n}(k(t))B_{nK}(t)
-F(t)\sum_{n'}R_{nn'}(k(t))B_{n'K}(t)\;,
\end{equation}
which is identical to Eq.(\ref{final}). The integration method in Section II
gives us linearly independent solutions
\begin{equation}
\psi_{Kj}^A(x,t)=\sum_{n}C_{Kjn}(t)\psi_{nK}(x,t)\;.
\end{equation}
Expanding the wavefunction $\psi(x,t)=\sum_{Kj}\alpha_{Kj}\psi_{Kj}^A(x,t)$,
we get
\begin{equation}
\rho_{nK}(t)=
\sum_{j'j''}\alpha_{Kj'}\alpha_{Kj''}^\ast
e^{i(\omega_{Kj'}-\omega_{Kj''})t}
P_{Kj'n}^0(t){P_{Kj''n}^0}^\ast(t)\;.
\end{equation}
The diagonal elements have a different appearance as compared to
Eq.(\ref{rhodiasca}), but the conclusion concerning the response frequency
spectrum is the same as before the gauge change.

\epsfxsize=14cm
\hspace{0.5cm}%
\epsfbox{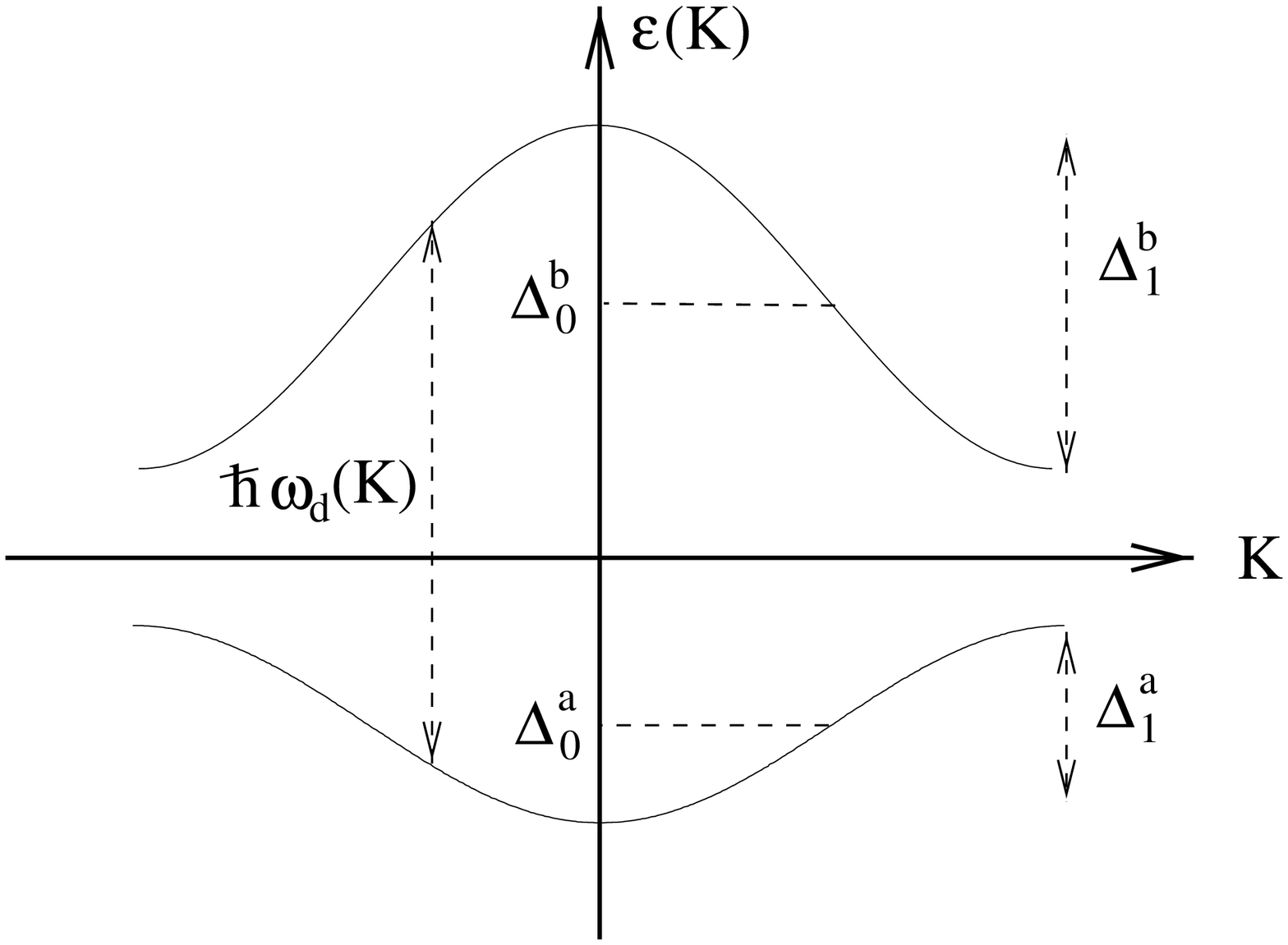}
\begin{figure}
\caption{Definitions of various energy parameters for the two-band
superlattice. $\Delta_0^{a,b}$ are the band mid-points, $\Delta_1^{a,b}$
their widths, and $\hbar\omega_d(K)=\epsilon_b(K)-\epsilon_a(K)$ is
the band separation at $K$.  Also in the text are used:
(i) $\Delta\omega_d(I)
={\rm max}[\omega_d(K)]-{\rm min}[\omega_d(K)],\;K\in I$;
(ii) $\widetilde{\omega}_d(K)=\omega_d(K) + FX_-/\hbar$ (here
$X_-=X^a-X^b$, where $X^i$ are defined in Eq.(\protect{\ref{tbhamilt}}));
(iii) $\langle\omega_d\rangle_I$, which is the average of $\omega_d(K)$,
over a given part of the Brillouin zone $I$;
(iv) $\omega_0=\langle \omega_d(K)\rangle_{BZ}$, i.e.,
the average band separation over the entire Brillouin zone, and, finally
(v) $\widetilde{\omega}_0=\omega_0+FX_-/\hbar$.}
\label{fig1}
\end{figure}

\newpage

\epsfxsize=14cm
\hspace{0.5cm}
\epsfbox{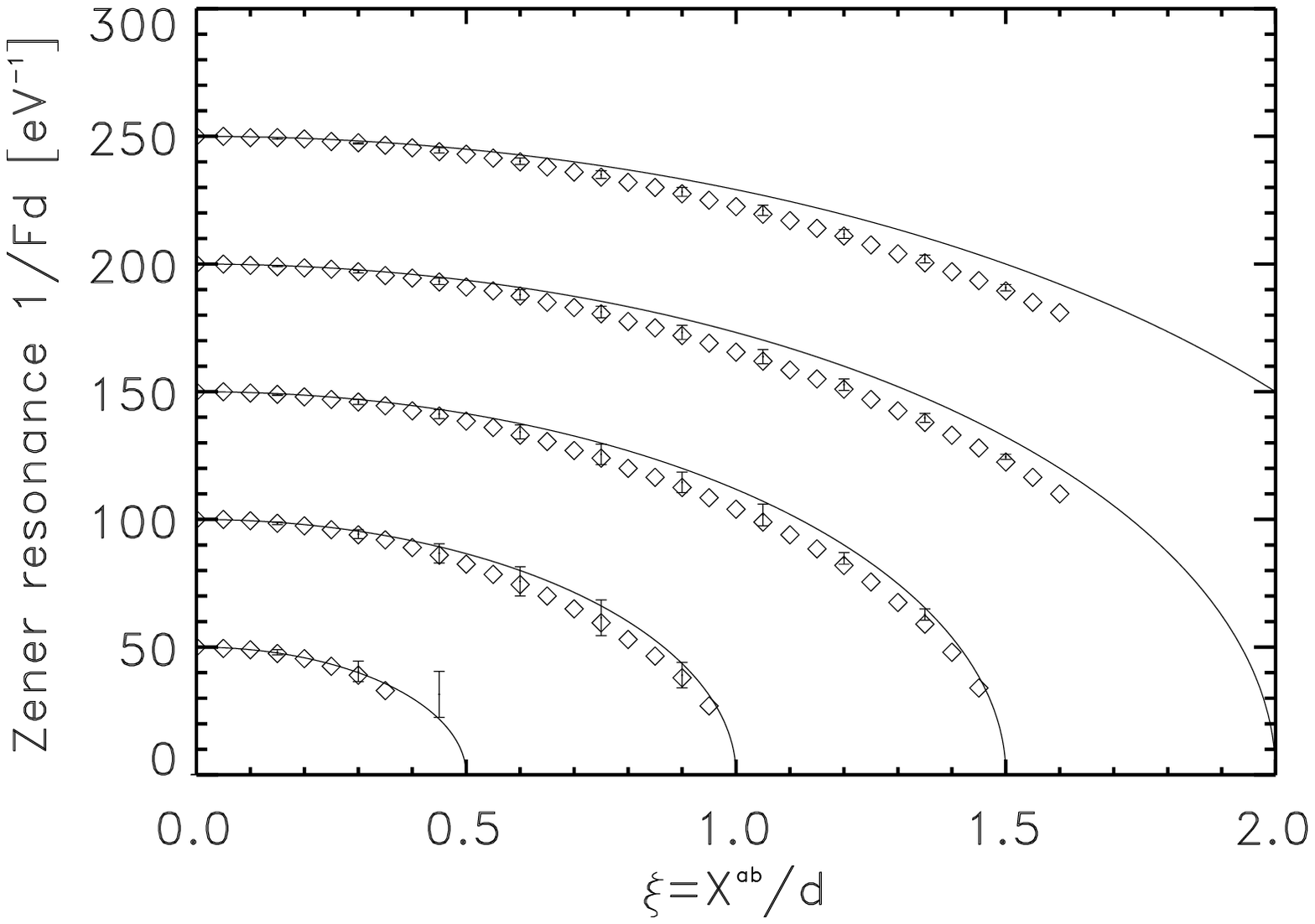}
\begin{figure}
\caption{The first five Zener resonances as a function of the band coupling
parameter
$\xi = X^{ab}/d$. The diamonds represent the field values at which the
distance between WS ladders has a local minimum. The solid lines give the
local minima for the flat band model. The 'error bars' indicate at which
field strengths the electronic response shows a Zener resonance.
The finite widths are a consequence of dispersion. See the text for further
discussion. The superlattice parameters are $X_-=0$, $d=10$nm,
$\Delta_1^a=14$meV, $\Delta_1^b=14$meV, and
$\Delta^{ab}\equiv\Delta_0^b-\Delta_0^a=20$meV.}
\label{fig2}
\end{figure}


\begin{thebibliography}{99}
\bibitem{Proof1}

This is seen by substituting
$t=\tilde{t}+T_0$,
$\tilde{E}(t)=E(t+T_0)$,
$\tilde{A}(t)=-\int_0^t \tilde{E}\,dt'$, and
$\widetilde{C}_{nK}(t)=C_{nK}(t+T_0)$.
Then $k(\tilde{t}+T_0)=K+eA(T_0)/\hbar+e\tilde{A}(\tilde{t})/\hbar$.
Using this in Eq.(\ref{final}) we get the original form with $K$
replaced by $K+eA(T_0)/\hbar$. Eq.(\ref{condperiodic}) becomes
$\widetilde{C}_{nK}(\tilde{t}+T_{ac})
=\widetilde{C}_{n,K+\Delta K}(\tilde{t})e^{-i\epsilon T_{ac}/\hbar}$.
\bibitem{Proof2}
The proof goes as follows.  A change
$R_{nn}(k)\rightarrow R_{nn}(k)+\Delta R(k)$ in Eq.(\ref{final}) is
compensated for by defining
$C_{nK}(t)=\widetilde{C}_{nK}(t)e^{\frac{i}{\hbar}\int_0^t F\Delta R\,dt'}$.
This brings Eq.(\ref{final}) back to the original form.
Eq.(\ref{dcdirectcond}) becomes
$\widetilde{C}_{nK}(t+T_B)
=\widetilde{C}_{nK}(t)e^{-i(\epsilon+F\langle\Delta R\rangle)T_B/\hbar}$ where
$\langle ... \rangle$ is the period average. We see that the WS ladders
are translated by the same amount $-F\langle\Delta R\rangle$.
(For a time-periodic field and a time-independent $\Delta R$
Eq.(\ref{condperiodic}) is replaced by
$\widetilde{C}_{nK}(t+T_{ac})=\widetilde{C}_{nK+\Delta K}(t)
e^{-i(\epsilon+\langle F\rangle\Delta R)T_{ac}/\hbar}$, thus the quasienergy
ladders are translated by $-\langle F\rangle\Delta R$.)
\end{thebibliography}
\end{document}